\documentclass{article}
\title{Weak Kleene Algebra is Sound and (Possibly) Complete for Simulation}
\author{Ernie Cohen\\Microsoft}
\usepackage{abbrev}
\usepackage{latexsym}
\usepackage{amsfonts}

\newcommand\Def[1]{{\em #1}}
\newcommand\Ref[1]{(\ref{#1})}

\newcommand\True{\mathit{true}}
\newcommand\So{\mbox{ so }}
\newcommand\Dist{\mbox{distr}}
\newcommand\Check[1]{{#1\!\downarrow}}
\newcommand\CheckIf[2]{{\Check{#1} \, #2}}
\newcommand\Tr[2]{{#1}'}
\newcommand\Trans[3]{{#1 \!\stackrel{#2}\rightarrow #3}}
\newcommand\Logic{\mbox{logic}}

\newcommand\Hyp{\mbox{(hyp)}}

\newcommand\AndSym{\land}

\newcommand\Defof{\mbox{def }}
\newcommand\ImpliesSym{\Rightarrow}
\newcommand\ImpliedSym{\Leftarrow}

\newcommand\If{\mbox{{\bf if }}}
\newcommand\Else{\mbox{{\bf else }}}
\newcommand\ByCheck[1]{\Check{#1}}
\newcommand\ByTr{\mbox{def of }\ \Tr{}{a}}
\newcommand\Induc{\mbox{induc hyp}}

\newenvironment{Proof}{
\[\begin{array}{l@{\hspace{5mm}}l@{\hspace{5mm}\{}l@{\}}}}{\end{array}\]}

\newcommand\Set[1]{\{#1\}}
\newcommand\EmptySet{\{\}}

\newcommand\V{\Box}

\begin{abbreviate}{Math}{:!(|<>/|.?-}{:!(|<>/.?-}
  \x ..   {\ }
  \x .    {.}
  \x :s:  {^*}
  \x :s   {^*}
  \x :    {:}
  \x (A:  {(\forall}
  \x (A   {(A}
  \x (E:  {(\exists\; }
  \x (E   {(E}
  \x (    {(}
  \x ->   {\ImpliesSym}
  \x ---  {---}
  \x --   {--}
  \x -    {-}
  \x <=>  {\Leftrightarrow}
  \x <=   {\ImpliedSym}
  \x <<<  {<}
  \x <<   {\preceq}
  \x <    {\leq}
  \x >>   {>}
  \x >    {\geq}
  \x |-   {\vdash}
  \x |    {|}
  \x /|   {\AndSym}
  \x /    {/}
  \x !    {\lnot}
  \x ?    {\lor}
  \x ~    {\sim}
\end{abbreviate}

\begin{document}
\maketitle
\begin{abstract}
We show that the axioms of Weak Kleene Algebra (WKA) are sound and
complete for the theory of regular expressions modulo simulation
equivalence, assuming their completeness for monodic trees (as
conjectured by Takai and Furusawa).
\end{abstract}

\Math

\section{Introduction}
Kleene algebra \cite{Kozen} completely axiomatizes (the equational
theory of) regular expressions modulo \Def{trace} equivalence. Here,
we show that \Def{Weak Kleene Algebra} (WKA) completely axiomatizes
regular expressions modulo \Def{simulation} equivalence, assuming that
they are complete for the equational theory of monodic trees. Their
completeness for monodic trees was claimed by Takai and Furusawa
\cite{MTKA}, but later errors in their proof were discovered and have
not yet been fixed.
 
By contrast, Horn axiomatization of regular expressions modulo \Def{bisimulation}
equivalence remains an open problem.

\section{Regular Expressions and Simulation}

Milner \cite{Milner} proposed the following process interpretation of regular
expressions, typically presented as a structured operational semantics
over the relations $\Check{x}$ (meaning that $x$ is in an
``accepting'' state) and $\Trans{x}{a}{y}$ (meaning that state $x$ can
evolve into state $y$ through an $a$-labeled transition), where $a$ is a
metavariable ranging over transition labels. We present the theory
using calculus notation: $\Tr{x}{a}$ is the set of terms $y$ such that
$\Trans{x}{a}{y}$. 
(The transition rules and proofs are uniform in the
transition label, so proofs are given only for $a$.) As usual, we omit the product operator
in writing regular expressions (i.e., we write $x \cdot y$ as $x..y$).
We extend the regular expression operators to sets of terms by
pointwise application, and cast terms to singleton sets when 
necessary. Finally, we extend $\ \Check{x}$ to a prefix operator on sets:
$\CheckIf{x}{S}$ is $S$ if $\Check{x}$, and is the empty set
otherwise. Using these conventions, Milner's semantics (as presented
in \cite{Baeten}) is as follows:

\[\begin{array}{l@{\hspace{1in}}rcl}
\neg (\Check{0})                            & \Tr{0}{a} &=& \EmptySet\\
\Check{1}                                   & \Tr{1}{a} &=&\EmptySet\\
\neg (\Check{b})                            & \Tr{b}{a} &=& \If\ (a=b)\ \Set{1}\ \Else\ \EmptySet\\
\Check{(x+y)} <=> \Check{x} ? \Check{y}     & \Tr{(x+y)}{a} &=& \Tr{x}{a} \cup \Tr{y}{b}\\
\Check{(x..y)} <=> \Check{x} /| \Check{y}   & \Tr{(x..y)}{a} &=& \Tr{x}{a}..y \cup \CheckIf{x}{\Tr{y}{a}}\\
\Check{x:s}                                 & \Tr{(x:s)}{a} &=&\Tr{x}{a}..x:s\\
\end{array}\]
(Note that  $\Tr{x}{a}$ is always finite (by structural induction on
$x$).) A binary relation $R$ is a \Def{simulation} iff
\[ (A:x,y: R(x,y) -> (\Check{x} -> \Check{y}) /| (A:a: R(x',y'))) \]
where $R$ has been extended to sets by 
\[ R(S,T)  <=> (A:x \in S: (E: y \in T: R(x, y)))\]
Define $<$ as the maximal 
simulation relation, and simulation equivalence $=$ by
$ x=y <=> x<y /| y<x $.
Note that simulation equivalence is weaker than bisimulation equivalence,
which requires the two simulations to be inverses. For example, $a..b
+ a..(b+c)$ and $a..(b+c)$ are simulation equivalent, but not
bisimulation equivalent. 

\section{Weak Kleene Algebra}
Weak Kleene algebra (WKA) is given by the following axioms over the language
of regular expressions (where $x<y$ abbreviates $x+y=y$).
\begin{eqnarray}
\label{PlusAssoc}&&|-(x+y)+z = x+(y+z)\\
\label{PlusIdem}&&|-x+x = x\\
\label{PlusComm}&&|- x+y = y+x\\
\label{PlusZero}&&|-x+0 = x\\
\label{OneSeq}&&|-1..x = x\\
\label{SeqOne}&&|-x..1 = x\\
\label{ZeroSeq}&&|-0..x = 0\\
\label{SeqAssoc}&&|-(x..y)..z = x..(y..z)\\
\label{PlusSeq} &&|-(x+y)..z = x..z+y..z\\
\label{SeqPlus}&&|-x..y+x..z < x..(y+z)\\
\label{StarExp}&&|-x:s = 1 + x..x:s\\
\label{LInduc} &&y..x<x |- y:s:x < x\\
\label{RInduc} &&x..(y+1) < x |- x..y:s < x
\end{eqnarray}
We write $|-x=y$ to mean that $x=y$ is provable from the WKA axioms.
It is easy to prove using these axioms that
$<$ is a partial order, and that the regular operators are monotonic wrt. $<$.
A number of similar process algebras and Kleene-like algebras
(e.g. \cite{LKA}) have been proposed, but lacking the weak right
induction law \Ref{RInduc}, which was first proposed in
\cite{CohenTalk99}. The WKA axioms were independently proposed for
probabilistic programs \cite{CohenTalk} and for monodic regular tree
languages \cite{MTKA}.

\section{Soundness}
We prove that the WKA axioms are sound for simulation ---
\begin{equation}
\label{soundness}
(|- x=y)\ ->\ x=y
\end{equation}
--- as follows. Given a set of axioms, take the set of instances of these
axioms whose hypotheses are satisfied
by $<$, and in each conclusion replace $x<y$ with $x R y$ and replace
$x=y$ with $x R y /|y R x$, where $R$ is a fresh binary relation symbol. Let $R$ be the weakest relation satisfied
by these conclusions; we prove $R\ \subseteq\ \mathalpha{<}$ (which implies that
$<$ is closed under the axioms, and hence the axioms are sound). Defining

\begin{eqnarray*}
\overline{R} &=& R\ \cup \mathalpha{<}\\
<<    &=&\mbox{the reflexive, transitive closure of }\overline{R}\\
x ~ y &<=>& x << y /| y << x \\
S << T &<=>& (A: x \in S: (E: y \in T: x << y))\\
S ~ T &<=>& S<<T /| T<<S
\end{eqnarray*}
we show  $R\ \subseteq\ <$ as follows:
\begin{Proof}
R  \subseteq \mathalpha{<} &<=&R \subseteq \mathalpha{<<}\\
\mathalpha{<<} \subseteq \mathalpha{<} &<=& \Defof <\\
<<\mbox{ is a simulation}&<=&\Defof \mbox{simulation}\\
(A: x,y: x<<y -> (\Check{x} -> \Check{y}) /| (A: a: x' << y')) &<=& \mbox{transitivity}\\
(A: x,y: x \overline{R} y -> (\Check{x} -> \Check{y}) /| (A: a: x' << y')) &<=&
  <\mbox{ a sim, }\mathalpha{<} \subseteq \mathalpha{<<}\\
(A: x,y: x R y -> (\Check{x} -> \Check{y}) /| (A: a: x' << y'))
\end{Proof}

Thus, to show soundness of $H |- x<y$ for simulation, we assume $H$ and
show $\Check{x} -> \Check{y}$ and $x' << y'$.
We do this (for all instances of) one axiom at a time, so we can use instances
of earlier axioms (since we have already proved that they are sound
for simulation).  Since $H |- x R y$ (by definition of $R$), 
we can also use instances of the axiom being proved, but
with $<$ and $=$ in the conclusion replaced with $<<$ and $~$.
Note also that while $'$ is $<$-monotonic (by the
definition of $<$), we cannot assume $'$ is $<<$-monotonic.

With this preparation, the soundness proofs (presented in section
\Ref{SoundnessAppendix}) are mostly routine, except for the
right-induction axiom \Ref{RInduc}, whose proof we give here
(even though it occurs logically after the others). 
This axiom is the most interesting because its soundness
depends on the finiteness of $\Tr{x}{a}$. To see why, suppose $x$ is
constructed as an infinite sum $x = (+n: (a+1)^n)$, and let $y$ be
$a$. Then $x..(y+1) < x$, but it is not the case that $x..y:s <
x$. ($x..y:s$ has infinite derivation chains, while $x$ does not.)

The proof of \Ref{RInduc} is as follows: assuming $x..(y+1) < x$,
\begin{Proof}
\Check{(x..y:s)} &->& \ByCheck{(u.v)}\\
\Check{x}
\end{Proof}
\begin{Proof}
\Tr{(x..y:s)}{a} &=&\Tr{(u..v)}{a}\\
\Tr{x}{a}..y:s \cup \CheckIf{x}{\Tr{y}{a}}..y:s &=&\Dist\\
(\Tr{x}{a} \cup \CheckIf{x}{\Tr{y}{a}})..y:s &<&\Tr{x}{a} = \Tr{x}{a}..1<\Tr{x}{a}..(y+1)\ \Ref{SeqOne},\Ref{SeqPlus}\\
(\Tr{x}{a}..(y+1) \cup \CheckIf{x}{\Tr{y}{a}})..y:s &=&\ByTr\\
\Tr{(x..(y+1))}{a}..y:s &<&x..(1+y)<x\ \Hyp;\ \mbox{monotonicity}\\
\Tr{x}{a}..y:s &<<&\mbox{proof below}\\
\Tr{x}{a}
\end{Proof}
To prove $\Tr{x}{a}..y:s <<\Tr{x}{a}$, define the binary relation $Q$ on (the set)
$\Tr{x}{a}$ by 
\[u Q v <=> (E: n >> 0: u..(y+1)^n<v)\]
Note $u Q v -> u<v /| u..(y+1)<v$ (by monotonicity), and that $Q$ is
total:
\begin{Proof}
\Tr{x}{a}..(y+1) &<&\Defof\ <\\
\Tr{x}{a}..(y+1) \cup \CheckIf{x}{\Tr{(y+1)}{a}} &=&\ByTr\\
\Tr{(x..(y+1))}{a} &<&x..(y+1) < x\ \Hyp\\
\Tr{x}{a}
\end{Proof}
Because $Q$ is a total, transitive relation on a finite set, for each
$u \in \Tr{x}{a}$, there is a $v$ such that $u Q v /| v Q v$; hence
$u..y:s << v$:
\begin{Proof}
u..y:s &<&u Q v,\ \So\ u<v\\
v..y:s &<<& v Q v,\ \So\ v..(y+1)<v;\ \Ref{RInduc}\\
v
\end{Proof}
Thus, for every $u \in \Tr{x}{a}$, $u..y:s << \Tr{x}{a}$, so
$\Tr{x}{a}..y:s << \Tr{x}{a}$. 

\section{Monodic Tree Languages}
\label{monodic}

A \Def{monodic tree language} is a set of first-order terms over a
first order language with a single variable $\V$.  For a term $t$ and
set of terms $S$, $t(S)$ is the set of terms obtained from $t$ by
replacing each variable instance with an element of $S$. (Thus, if $t$
has $n$ variable instances and $S$ is finite, $t(S)$ is a set of $\vert
S \vert^n$ terms.) Extend this to sets by 
$T(S) = (\cup_{t \in T}  t(S))$. 

An \Def{interpretation} (notation: $I,J$) is a function from regular
expression symbols to tree languages. Extend interpretations to
regular expressions as follows (when applied to tree languages, $=$
denotes ordinary set equality):
\begin{eqnarray*}
I(0) &=& \EmptySet\\
I(1) &=& \Set{\V}\\
I(x..y) &=& I(x)(I(y))\\
I(x+y) &=& I(x) \cup I(y)\\
I(x^*) &=& \cup_{n\geq 0} I((1+x)^n) \\
\end{eqnarray*}
In \cite{MTKA}, Takai and Furusawa prove that the 
WKA axioms are sound and complete for
the equational theory of regular expressions interpreted as tree
languages:
\begin{equation}
\label{mtCompleteness}
(|- x=y) <=> (A: I: I(x) = I(y))
\end{equation}

\section{Interpretation Respects Simulation}
Our last observation is that interpretation respects simulation equivalence:
\begin{equation} \label{InterpretationRespectsSimulation}
x=y -> I(x) = I(y)
\end{equation}
The proof depends on the following property of interpretations (proved
in section \Ref{InterpretationNormal} by induction on $x$):
\begin{equation} \label{IntNormal}
 I(x) = \CheckIf{x}{\Set{\V}} 
      \cup (\cup_{a,z\in \Tr{x}{a}}  I(a)(I(z)))
\end{equation}
To prove \Ref{InterpretationRespectsSimulation}, 
we first switch to an interpretation $J$ where $(A: a: \V \not\in J(a))$.
Let $f$ be a fresh unary function symbol, and define $J$ by
\[ J(a) = I(a) - \Set{\V} \cup \Set{f(\V)\ \vert\ \V\in I(a)} \]
For any regular expression $z$, it is
easy to show that $I(z)$ can be computed from $J(z)$ 
by repeatedly replacing $f(t)$ with $t$. 
Thus, it suffices to prove $x=y -> J(x)=J(y)$. 

We prove this by
proving the more general $(A: t,x,y:  t\in J(x)-J(y) ->  x\not<y)$, by
induction on $\vert t \vert$. For $t = \V$,
\begin{Proof}
\V \in J(x) &->&\Ref{IntNormal}\\
\Check{x} ? (E: a,u: u \in \Tr{x}{a} /| \V \in J(a)(J(u)))  &->& \V\notin J(a)\\
\Check{x} &->& \neg J(y),\ \Ref{IntNormal}\\
\Check{x} /| \neg \Check{y} &->& \Defof\ <\\
x\not< y
\end{Proof}
For $t \neq \V$, 
\begin{Proof}
t \in J(x)-J(y) &->& \Ref{IntNormal},\ t\neq\V\\
t \in (\cup_{a,z\in \Tr{x}{a}} J(a)(J(z)))  /| t\notin (\cup_{a,z \in \Tr{y}{a}}  J(a)(J(z))) &->&\cup\\
\multicolumn{3}{l}{(E:a,u: u \in \Tr{x}{a} /| t\in J(a)(J(u))  /| (A: v \in \Tr{y}{a}: t\notin J(a)(J(v))))}\\
     &->&\V \notin J(a)\\
\multicolumn{3}{l}{(E:a,u,t0:  \vert t0\vert <<< \vert t \vert /| u\in\Tr{x}{a} /|  t0\in J(u) 
  /| (A: v \in \Tr{y}{a}: t0 \notin J(v)))}\\
     &->&\Induc\\
(E:a,u: u \in \Tr{x}{A} /| (A: v \in \Tr{y}{a}: u\not< v)) &->&\Defof\ <\\
\Tr{x}{a}\not<\Tr{y}{a} &->&\Defof\ <\\
x\not<y
\end{Proof}

\section{Completeness}

The completeness of WKA for simulation 
\begin{equation}
x = y\ ->\ |- x = y
\end{equation}
now follows:
\begin{Proof}
x = y &->& \Ref{InterpretationRespectsSimulation}\\
(A: I: I(x) = I(y)) &->&\Ref{mtCompleteness}\\
|- x = y 
\end{Proof}



\section{Acknowledgements}
Bernhard M\"oller provided useful feedback on the model theory for WKA. 
Georg Struth first suggested looking into \cite{MTKA}. 
Annabelle McIver first proposed using the weak right induction law
\Ref{RInduc}. An anonymous referee found a number of typos, minor errors, 
and a significant error in the original soundness proof of \Ref{LInduc}.

\appendix
\section{Soundness of WKA for Simulation}
\label{SoundnessAppendix}

\begin{Proof}
\Ref{PlusAssoc}:\\
\Check{(x+(y+z))} &<=>&\ByCheck{(u+v)}\\
\Check{x} ? \Check{y} ? \Check{z} &<=>&\ByCheck{(u+v)}\\
\Check{((x+y)+z)}
\end{Proof}
\begin{Proof}
\Tr{(x+(y+z))}{a}&=&\Tr{(u+v)}{a}\\
\Tr{x}{a} \cup \Tr{y}{a} \cup \Tr{z}{a} &=&\Tr{(u+v)}{a}\\
\Tr{((x+y)+z)}{a}
\end{Proof}

\begin{Proof}
\Ref{PlusIdem}:\\
\Check{(x+x)}&<=>&\Tr{(u+v)}{a}\\
\Check{x}
\end{Proof}
\begin{Proof}
\Tr{(x+x)}{a}&=&\Tr{(u+v)}{a}\\
\Tr{x}{a}
\end{Proof}

\begin{Proof}
\Ref{PlusComm}:\\
\Check{(x+y)}&<=>&\ByCheck{(u+v)}\\
\Check{x} ? \Check{y}&<=>&\ByCheck{(u+v)}\\
\Check{(y+x)}
\end{Proof}
\begin{Proof}
\Tr{(x+y)}{a}&=&\Tr{(u+v)}{a}\\
\Tr{x}{a} \cup \Tr{y}{a} &=&\Tr{(u+v)}{a}\\
\Tr{(y+x)}{a}
\end{Proof}

\begin{Proof}
\Ref{PlusZero}:\\
\Check{(x+0)} &<=>&\ByCheck{(u+v)}\\
\Check{x} ? \Check{0} &<=>&\neg(\Check{0})\\
\Check{x}
\end{Proof}
\begin{Proof}
\Tr{(x+0)}{a} &=&\Tr{(u+v)}{a}\\
\Tr{x}{a} \cup \Tr{0}{a} &=&\Tr{0}{a}\\
\Tr{x}{a}
\end{Proof}

\begin{Proof}
\Ref{OneSeq}:\\
\Check{(1..x)}&<=>&\ByCheck{(u..v)},\ \Check{1}\\
\Check{x}
\end{Proof}
\begin{Proof}
\Tr{(1..x)}{a}&=&\Tr{(u..v)}{a},\ \Tr{1}{a},\ \ByCheck{1}\\
\Tr{x}{a}
\end{Proof}

\begin{Proof}
\Ref{SeqOne}:\\
\Check{(x..1)}&<=>&\ByCheck{(u..v)},\ \ByCheck{1}\\
\Check{x}
\end{Proof}
\begin{Proof}
\Tr{(x..1)}{a}&=&\Tr{(u..v)}{a},\ \Tr{1}{a}\\
\Tr{x}{a}..1 &~&\Ref{SeqOne}\\
\Tr{x}{a}
\end{Proof}

\begin{Proof}
\Ref{ZeroSeq}:\\
\Check{(0..x)} &<=>&\ByCheck{(u..v)}\\
\Check{0} /| \Check{x} &<=>&\neg(\Check{0})\\
\Check{0}
\end{Proof}
\begin{Proof}
\Tr{(0..x)}{a} &=&\Tr{(u..v)}{a}\\
\Tr{0}{a}..x \cup \Check{0}{\Tr{x}{a}} &=&\Tr{0}{a},\ \neg(\Check{0})\\
\EmptySet &=&\Tr{0}{a}\\
\Tr{0}{a}
\end{Proof}

\begin{Proof}
\Ref{SeqAssoc}:\\
\Check{((x..y)..z)}&<=>&\ByCheck{(u..v)}\\
\Check{x} /| \Check{y} /| \Check{z}&<=>&\ByCheck{(u..v)}\\
\Check{(x..(y..z))}
\end{Proof}
\begin{Proof}
\Tr{((x..y)..z)}{a}&=&\Tr{(u..v)}{a}\\
\Tr{(x..y)}{a}..z \cup \CheckIf{(x..y)}{\Tr{z}{a}}&=&\Tr{(u..v)}{a},\ \ByCheck{(u..v)}\\
(\Tr{x}{a}..y \cup \CheckIf{x}{\Tr{y}{a}})..z \cup \CheckIf{x}{\CheckIf{y}{\Tr{z}{a}}} &=&\Dist\\
(\Tr{x}{a}..y)..z \cup \CheckIf{x}{(\Tr{y}{a}..z \cup \CheckIf{y}{\Tr{z}{a}})}&~&\Ref{SeqAssoc}\\
\Tr{x}{a}..(y..z) \cup \CheckIf{x}{(\Tr{y}{a}..z \cup \CheckIf{y}{\Tr{z}{a}})}&=&\Tr{(u..v)}{a}\\
\Tr{x}{a}..(y..z) \cup \CheckIf{x}{\Tr{(y..z)}{a}}&=&\Tr{(u..v)}{a}\\
\Tr{(x..(y..z))}{a}
\end{Proof}

\begin{Proof}
\Ref{PlusSeq}:\\
\Check{((x+y)..z)} &<=>&\ByCheck{(u..v)},\ \Check{(u+v)}\\
(\Check{x} ? \Check{y}) /| \Check{z} &<=>&\Logic\\
(\Check{x} /| \Check{z}) ? (\Check{y} /| \Check{z}) &<=>&\ByCheck{(u..v)},\ \Check{(u+v)}\\
\Check{(x..z+y..z)}
\end{Proof}
\begin{Proof}
\Tr{((x+y)..z)}{a} &=& \Tr{(u..v)}{a}\\
\Tr{(x+y)}{a}..z \cup \CheckIf{(x+y)}{\Tr{z}{a}} &=&\Tr{(u+v)}{a},\ \ByCheck{(u+v)},\ \cup\\
(\Tr{x}{a} \cup \Tr{y}{a})..z \cup \CheckIf{x}{\Tr{z}{a}} \cup \CheckIf{y}{\Tr{z}{a}} &=&\Dist\\
(\Tr{x}{a}..z \cup \CheckIf{x}{\Tr{z}{a}}) \cup (\Tr{y}{a}..z \cup \CheckIf{y}{\Tr{z}{a}})  &=&\Tr{(u..v)}{a}\\
\Tr{(x..z)}{a} \cup \Tr{(y..z)}{a}&=&\Tr{(u+v)}{a}\\
\Tr{(x..z + y..z)}{a}
\end{Proof}

\begin{Proof}
\Ref{SeqPlus}:\\
\Check{(x..y+x..z)}&<=>&\ByCheck{(u+v)},\ \Check{(u..v)}\\
(\Check{x} /| \Check{y}) ? (\Check{x} /| \Check{z})&<=>& \Logic\\
\Check{x} /| (\Check{y} ? \Check{z})&<=>&\ByCheck{(u+v)},\ \Check{(u..v)}\\
\Check{(x..(y+z))}
\end{Proof}
\begin{Proof}
\Tr{(x..y + x..z)}{a} &=&\Tr{(u:s)}{a}\\
\Tr{x}{a}..y \cup \CheckIf{x}{\Tr{y}{a}} \cup \Tr{x}{a}..z \cup \CheckIf{x}{\Tr{z}{a}} &=&\cup\\
\Tr{x}{a}..y \cup \Tr{x}{a}..z \cup \CheckIf{x}{\Tr{y}{a}}  \cup \CheckIf{x}{\Tr{z}{a}} &=&\Dist\\
\Tr{x}{a}..y \cup \Tr{x}{a}..z \cup \CheckIf{x}{\Tr{(y+z)}{a}} &<<&\Ref{SeqPlus}\\
\Tr{x}{a}..(y+z) \cup \CheckIf{x}{\Tr{(y+z)}{a}} &=&\Tr{(u..v)}{a}\\
\Tr{(x..(y+z))}{a}
\end{Proof}

\begin{Proof}
\Ref{StarExp}:\\
\Check{x:s} &<=>&\ByCheck{u:s}\\
\True &<=>&\ByCheck{(u+v)},\ \Check{1}\\
\Check{(1 + x..x:s)}
\end{Proof}
\begin{Proof}
\Tr{x:s}{a} &=&u = \CheckIf{v}{u} \subseteq u\\
\Tr{x:s}{a} \cup \CheckIf{x}{\Tr{x:s}{a}} &=&\Tr{u:s}{a}\\
\Tr{x}{a}..x:s \cup \CheckIf{x}{\Tr{x:s}{a}} &=&\Tr{(u..v)}{a}\\
\Tr{(x..x:s)}{a} &=&\Tr{1}{a}\\
\Tr{1}{a} \cup \Tr{(x..x:s)}{a} &=&\Tr{(u+v)}{a}\\
\Tr{(1+x..x:s)}{a}
\end{Proof}

\begin{Proof}
\multicolumn{3}{l}{\Ref{LInduc}:\ \mbox{We prove soundness of the stronger axiom}}
\end{Proof}
\[\Ref{LInduc}'\ \ y..x < x |- z..(y:s:x) < z..x \]
\begin{Proof}
\multicolumn{3}{l}{\mbox{Assuming } y..x < x,}\\
\Check{(z..(y:s:x))} &<=>&\ByCheck{(u..v)},\ \Check{u:s}\\
\Check{z} /| \Check{x} &<=>&\ByCheck{(u..v)}\\
\Check{(z..x)}
\end{Proof}
\begin{Proof}
\Tr{(z..(y:s:x))}{a} &=&\Tr{(u..v)}{a},\ \Dist\\
\Tr{z}{a}..(y:s:x) \cup \Check{z}..\Tr{y:s}{a}..x \cup
  \Check{z}..\Check{y:s}..\Tr{x}{a} &=&\Tr{u:s}{a},\ \Check{u:s}\\
\Tr{z}{a}..(y:s:x) \cup \Check{z}..(\Tr{y}{a}..y:s)..x \cup \Check{z}..\Tr{x}{a} &<&\Ref{SeqAssoc}\\
\Tr{z}{a}..(y:s:x) \cup \Check{z}..\Tr{y}{a}..(y:s..x) \cup \Check{z}..\Tr{x}{a} &<<&y..x < x;\ \Ref{LInduc}'\\
\Tr{z}{a}..x \cup \Check{z}..\Tr{y}{a}..x \cup \Check{z}..\Tr{x}{a} &<& y..x<x,\ \So\ \Tr{y}{a}..x < \Tr{x}{a}\\
\Tr{z}{a}..x \cup \Check{z}..\Tr{x}{a} &=& \Tr{(u..v)}{a}\\
\Tr{(z..x)}{a}
\end{Proof}

\section{Interpretations as Trees}
\label{InterpretationNormal}
Here, we prove \Ref{IntNormal}:
\begin{eqnarray*}
 I(x) = \CheckIf{x}{\Set{\V}} \cup (\cup_a I_a(x))
\end{eqnarray*}
where
\[ I_a(x) = (\cup_{z \in \Tr{x}{a}} I(a)(I(z))) \]

We prove this by induction on the structure of $x$.

\begin{Proof}
x = 0:\\
I(0) &=& \Defof\ I\\
\EmptySet &=&\neg (\Check{0})\\
\CheckIf{0}{\Set{\V}} &=& \Tr{0}{a} = \EmptySet,\ \So\ I_a(0) = \EmptySet\\
\CheckIf{0}{\Set{\V}} \cup (\cup_a I_a(0)) 
\end{Proof}

\begin{Proof}
x = 1:\\
I(1) &=& \Defof\ I\\
\Set{\V} &=&\Check{1}\\
\CheckIf{1}{\Set{\V}} &=& \Tr{1}{a} = \EmptySet,\ \So\ I_a(0) = \EmptySet\\
\CheckIf{1}{\Set{\V}} \cup (\cup_a I_a(1)) 
\end{Proof}

\begin{Proof}
x = b:\\
I(b) &=&\Defof\ ()\\
I(b)(\Set{\V}) &=&\cup\\
(\cup_{a=b} I(a)(\Set{\V})) &=&a=b -> \Tr{b}{a} = \Set{\V}\\
(\cup_{a=b} I(a)(\Tr{b}{a})) &=&\cup\\
(\cup_{a=b} (\cup_{z\in\Tr{b}{a}} I(a)(I(z))))&=&a \neq b -> \Tr{b}{a} = \EmptySet \\
(\cup_a (\cup_{z\in\Tr{b}{a}} I(a)(I(z))))&=&\Defof\ I_a\\
(\cup_a I_a(b)) &=&\neg(\Check{b})\\
\CheckIf{b}{\Set{\V}} \cup (\cup_a I_a(b))
\end{Proof}

\begin{Proof}
x = u+v:\\
I(u+v) &=&\Defof\ I\\
I(u) \cup I(v)&=&\Induc\\
(\CheckIf{u}{\Set{\V}} \cup (\cup_a I_a(u))) \cup (\CheckIf{v}{\Set{\V}} \cup (\cup_a I_a(v))) &=&\cup\\
(\CheckIf{u}{\Set{\V}} \cup \CheckIf{v}{\Set{\V}}) \cup ((\cup_a I_a(u)) \cup (\cup_a I_a(v))) &=&\Dist,\ \ByCheck{(x+y)}\\
\CheckIf{(u+v)}{\Set{\V}} \cup (\cup_a I_a(u)) \cup (\cup_a I_a(v))&=&\cup\\
\CheckIf{(u+v)}{\Set{\V}} \cup (\cup_a (I_a(u) \cup I_a(v))) &=&\Defof\ I_a\\
\CheckIf{(u+v)}{\Set{\V}} \cup (\cup_a (\cup_{z \in \Tr{u}{a} \cup \Tr{v}{a}} I(a)(I(z))))  &=& \Tr{(x+y)}{a}\\
\CheckIf{(u+v)}{\Set{\V}} \cup (\cup_a (\cup_{z \in \Tr{(u+v)}{a}} I(a)(I(z))))  &=& \Defof\ I_a\\
\CheckIf{(u+v)}{\Set{\V}} \cup (\cup_a I_a(u+v))
\end{Proof}

\begin{Proof}
x = u..v:\\
I(u..v) &=& \Defof\ I\\
I(u)(I(v)) &=& \Induc\\
(\CheckIf{u}{\Set{\V}} \cup (\cup_a I_a(u)))(I(v)) &=&\Dist\\
\CheckIf{u}{\Set{\V}}(I(v)) \cup (\cup_a I_a(u))(I(v))  &=&\V(S) = S\\
\CheckIf{u}{I(v)} \cup (\cup_a I_a(u))(I(v))  &=&\Induc\\
\CheckIf{u}{(\CheckIf{v}{\Set{\V}} \cup (\cup_a I_a(v)))} \cup (\cup_a I_a(u))(I(v))  &=&\Dist\\
\CheckIf{u}{\CheckIf{v}{\Set{\V}}} \cup (\cup_a I_a(u))(I(v)) \cup \CheckIf{u}{(\cup_a I_a(v))} &=&\ByCheck{(x..y)}\\
\CheckIf{(u..v)}{\Set{\V}} \cup (\cup_a I_a(u))(I(v)) \cup \CheckIf{u}{(\cup_a I_a(v))} &=&\mbox{(a) below}\\
\CheckIf{(u..v)}{\Set{\V}} \cup (\cup_a I_a(u..v))
\end{Proof}

\begin{Proof}
\mbox{(a):}\\
(\cup_a I_a(u))(I(v)) \cup \CheckIf{u}{(\cup_a I_a(v))} &=&\cup,\ \Defof\ I_a \\
(\cup_a (\cup_{z \in \Tr{u}{a}} I(a)(I(z))(I(v)))) \cup \CheckIf{u}{(\cup_a I_a(v))} &=&() \mbox{ assoc}\\
(\cup_a (\cup_{z \in \Tr{u}{a}} I(a)(I(z)(I(v))))) \cup \CheckIf{u}{(\cup_a I_a(v))} &=&I(x..y)\\
(\cup_a (\cup_{z \in \Tr{u}{a}} I(a)(I(z..v)))) \cup \CheckIf{u}{(\cup_a I_a(v))} &=&\cup\\
(\cup_a (\cup_{z \in (\Tr{u}{a}\,v)} I(a)(I(z)))) \cup \CheckIf{u}{(\cup_a I_a(v))} &=&\Dist\\
(\cup_a (\cup_{z \in (\Tr{u}{a}\,v)} I(a)(I(z)))) \cup (\cup_a \CheckIf{u}{I_a(v)}) &=&\Dist\\
(\cup_a (\cup_{z \in (\Tr{u}{a}\,v)} I(a)(I(z))) \cup \CheckIf{u}{I_a(v)}) &=&\Defof\ I_a\\
(\cup_a (\cup_{z \in (\Tr{u}{a}\,v)} I(a)(I(z)))) \cup \CheckIf{u}{(\cup_{z\in \Tr{v}{a}} I(a)(I(z)))} &=&\Dist\\
(\cup_a (\cup_{z \in (\Tr{u}{a}\,v)} I(a)(I(z)))) \cup (\cup_{z\in \CheckIf{u}{\Tr{v}{a}}} I(a)(I(z))) &=&\Dist\\
(\cup_a (\cup_{z \in (\Tr{u}{a}\,v \cup \CheckIf{u}{\Tr{v}{a}})} I(a)(I(z)))) &=&\Tr{(x..y)}{a}\\
(\cup_a (\cup_{z \in \Tr{(u..v)}{a}} I(a)(I(z)))) &=&\Defof\ I_a\\
(\cup_a I_a(u..v))
\end{Proof}

\begin{Proof}
\multicolumn{3}{l}{\mbox{Let $n$ and $m$ range over natural numbers;
    defining $U_m = I((1+u)^n)$,}}\\
I(u:s) &=&I(x:s)\\
(\cup_n U_n)) &=&\mbox{(a) below}\\
(\cup_n (\Set{\V} \cup (\cup_a I_a(u))(\cup_{m<<<n} U_m))) &=&\Dist\\
\Set{\V} \cup (\cup_n (\cup_a I_a(u))(\cup_{m<<<n} U_m)) &=&\mbox{(b),(a) below}\\
\Set{\V} \cup (\cup_a I_a(u))(\cup_n (\cup_{m<<<n} U_m)) &=&\cup\\
\Set{\V} \cup (\cup_a I_a(u))(\cup_n U_n) &=&I(x:s)\\
\Set{\V} \cup (\cup_a I_a(u))(I(u:s)) &=&\ByCheck{x:s}\\
\CheckIf{u:s}{\Set{\V}}\cup (\cup_a I_a(u))(I(u:s)) &=&\Defof\ I_a\\
\CheckIf{u:s}{\Set{\V}}\cup (\cup_a (\cup_{z\in \Tr{u}{a}} I(a)(I(z))))(I(u:s)) &=&\Dist\\
\CheckIf{u:s}{\Set{\V}}\cup (\cup_a (\cup_{z\in \Tr{u}{a}} (I(a)(I(z)))(I(u:s)))) &=&\mbox{assoc}\\
\CheckIf{u:s}{\Set{\V}}\cup (\cup_a (\cup_{z\in \Tr{u}{a}} I(a)(I(z)(I(u:s))))) &=&I(x..y)\\
\CheckIf{u:s}{\Set{\V}}\cup (\cup_a (\cup_{z\in \Tr{u}{a}} I(a)(I(z..u:s)))) &=&\cup\\
\CheckIf{u:s}{\Set{\V}}\cup (\cup_a (\cup_{z\in (\Tr{u}{a}\,u:s)} I(a)(I(z)))) &=&\Tr{x:s}{a}\\
\CheckIf{u:s}{\Set{\V}}\cup (\cup_a (\cup_{z\in \Tr{u:s}{a}} I(a)(I(z)))) &=&\Defof\ I_a\\
\CheckIf{u:s}{\Set{\V}}\cup (\cup_a I_a(u:s))
\end{Proof}

\begin{Proof}
\multicolumn{3}{l}{\mbox{(a) } U_n = \Set{\V} \cup (\cup_a
  I_a(u))(\cup_{m<<<n} U_m)} \\
\multicolumn{3}{l}{\mbox{Proof by induction on $n$. For $n=0$, this
  follows from $I(1)$; for $n >> 0$,}}\\
U_{n+1} &=&\mbox{(c) below}\\
U_n \cup (\cup_a I_a(u))(U_n) &=&\Induc\\
\Set{\V} \cup (\cup_a I_a(u))(\cup_{m<<<n} U_m) \cup (\cup_a
  I_a(u))(U_n) &=&U_m \subseteq U_n \mbox{ (c)}\\
\Set{\V} \cup (\cup_a I_a(u))(\cup_{m<<<n+1} U_m)
\end{Proof}

\begin{Proof}
\mbox{(b): For $\cup$-increasing sequence of languages $T_n$,}\\
(\cup_{n\geq 0} L(T_n)) = L(\cup_{n \geq 0} T_n) \\
\mbox{(because each tree has a finite number of $\V$'s)}
\end{Proof}

\begin{Proof}
\multicolumn{3}{l}{\mbox{(c) } U_{n+1} = U_n \cup (\cup_a I_a(u))(U_n) }\\
U_{n+1} &=&x^{n+1} = x..x^n\\
I((1+u)(1+u)^n) &=&I(x..y)\\
I(1+u)(U_n) &=&I(x+y)\\
(I(1)+I(u))(U_n) &=&I(1)\\
(\Set{\V}+I(u))(U_n) &=&\Induc\mbox{ (above)}\\
(\Set{\V} \cup (\CheckIf{u}{\Set{\V}} \cup (\cup_a I_a(u))))(U_n) &=&\CheckIf{u}{\Set{\V}} \subseteq\Set{\V}\\
(\Set{\V} \cup (\cup_a I_a(u)))(U_n) &=&\cup,\ \Defof\ \V()\\
U_n \cup (\cup_a I_a(u))(U_n) 
\end{Proof}

\end{document}